\documentstyle[epsfig,twocolumn,aps]{revtex}
\begin{document}
\draft
\title{Flat histogram simulation of lattice polymer systems}
\author{Lik Wee Lee and Jian-Sheng Wang}
\address{Department of Computational Science,
	National University of Singapore,
	Singapore 119260, Republic of Singapore}
\date{25 June 2001}
\maketitle
\begin{abstract}
We demonstrate the use of a new algorithm called the Flat Histogram sampling
algorithm for the simulation of lattice polymer systems. Thermodynamics
properties, such as average energy or entropy and other physical quantities
such as end-to-end distance or radius of gyration can be easily calculated
using this method. Ground-state energy can also be determined. We also explore
the accuracy and limitations of this method.
\end{abstract}
\pacs{02.70.Uu, 05.10.Ln, 75.40.Mg}

\narrowtext

\section{Introduction}
With the rapid rise of computing power, Monte Carlo methods \cite{Binder}
have become an important tool for studying high-dimensional systems such as
proteins, polymers and spin-glass models where many questions remain to be
answered.
While the Metropolis algorithm \cite{Metropolis}, due to its simplicity,
remains the most popular choice of method, it faces some severe drawbacks.
Firstly, the dynamics are slow for a class of problems that involve rugged
energy landscapes with multiple local minima. Secondly, a series of simulation
at different temperatures is needed to obtain the temperature dependence of
thermodynamic quantities. Coupled with slow dynamics, the computation time
can be prohibitive. Thirdly, it is difficult to calculate free energy or
entropy using this method.

While the use of histogram method \cite{Ferrenberg} can alleviate the
second problem by reweighting the canonical distribution, the accuracy is
limited to a small region in a parameter space. Recent methods based
on the direct computation of the density of states are capable of overcoming
the abovementioned drawbacks of Metropolis algorithm. The multicanonical 
method \cite{Berg} is the earliest of these. Entropic sampling \cite{Lee}
is commonly cited as an equivalent but simpler formulation of multicanonical
method. The flat histogram sampling algorithm \cite{Wang-broad} is able to
generate a flat histogram in energy space similar to multicanonical
simulations, but in a simpler and more efficient way. The transition matrix
Monte Carlo method \cite{Wang-tmc,Wang-tmmc,Li-thesis} can be used to either
obtain the density of states directly or construct the canonical distribution
for different temperatures. The method is based upon the definition of a
stochastic matrix, the transition matrix $T(E \to E')$. The flat histogram
sampling algorithm is an ideal algorithm for obtaining the transition matrix
elements through simulations.

The transition matrix Monte Carlo method and associated flat histogram sampling
algorithm are closely related to the broad histogram method 
\cite{Oliveira,Oliveira-EurB}. However, when it was first proposed by Oliveira
{\it et al} in 1996, the dynamics gave incorrect results in that the method did
not generate true microcanonical averages \cite{Wang-broad}. When corrected,
the flip rate is identical to flat histogram's but the name remains a historic
misnomer. A central quantity in the broad histogram formulation is
$\langle N(\sigma,\Delta E)\rangle_E$, the microcanonical
average of the number of potential moves that increase energy by $\Delta E
= E' - E$.
Using this definition together with the requirement that moves must be
reversible, a general equation called the broad histogram equation can be
derived \cite{Oliveira-exact}.

Although the definition of $\langle N(\sigma,\Delta E)\rangle_E$ works well
for Ising model and can be used to construct the transition matrix
$T(E \to E')$, this interpretation is less general and poses a problem for
other class of problems, such as polymer system. We shall define a more general
quantity, $T_\infty(E \to E')$,
the transition matrix at infinite temperature or simply called ``infinite
temperature transition matrix''.  The matrix elements of the infinite
temperature transition matrix reduces to $\langle N(\sigma, \Delta E)\rangle_E$
for some particular cases. The density of states $n(E)$, corresponds to the
left eigenvector of the infinite temperature transition matrix. However, it is
easier to obtain the density of states through another set of equations, the
detailed balance equations (analogous to the broad histogram equation)
explained later. Our procedure for solving the density of states is also
different from what is prescribed by the broad histogram method.

In the following section, we shall briefly describe our simulation model, the
HP model and its connection to protein folding. The subsequent section is
on the transition matrix Monte Carlo and flat histogram sampling algorithm. We
shall discuss how it can be applied to polymer models. We give some numerical
results in the fourth section and the conclusion in fifth section.

\section{The HP Model}

One of the most challenging problems in computational biology is the problem
of protein folding. Proteins are heteropolymer consisting of long chains
of amino acids. It is observed that proteins generally adopt a single
unique ``native'' structure. The biological function of a protein is often
intimately dependent upon the precise geometrical structure of this folded
native state. How does the protein encodes this unique native state in an
extremely large conformational space? Understanding this will be a major
breakthrough with implications in biochemistry and drug design.

It is believed that the native state lies at the global free energy
minimum. Only a small fraction of the total conformational space can be explored
using high-resolution force-field methods. Even more computing power is
required when solvent effects are included. Hence coarse grained models 
have been developed to model the folding process. The HP model \cite{Lau-HP}
is a commonly used simple lattice model with the basic assumption
that it is the hydrophobic forces that determines the native structure of the
protein. The model recognizes only two types of amino acids:
hydrophobic ($\mathsf{H}$) and polar ($\mathsf{P}$).
There are some good arguments for supporting this assumption
which we shall not elaborate \cite{Dill-review}.
Our chief concern is to study the statistical mechanical aspect of the model
and the performance of our algorithm. Given a sequence of $\mathsf{H}$ and
$\mathsf{P}$, each self-avoiding chain on a two-dimensional lattice is counted
as one conformation. Only non-bonded neighbouring $\mathsf{HH}$ contributes to
the total energy, i.e. $\epsilon_{\mathsf{HH}} = -1$ and
$\epsilon_{\mathsf{HP}} = \epsilon_{\mathsf{PP}} = 0$. Under
such conditions, low-energy conformations are compact with $\mathsf{H}$
residues residing mainly in the core and $\mathsf{P}$ residues on the outside.
The principle disadvantage of this model is that it leads to highly degenerate
ground states, especially in three dimensions \cite{Clote,Yue}.

\section{Transition Matrix Monte Carlo and Flat Histogram Sampling Method}

In a Monte Carlo simulation, it is usual to generate a sequence of states
$\{\sigma^1, \sigma^2, \ldots\}$ using a Markov chain where each
state denoted by $\sigma$ lies  in the phase space of the model.
The Markov chain is defined by a transition matrix
$W(\sigma\to\sigma')$. This stochastic matrix must satisfy
$\sum_{\sigma'} W(\sigma\to\sigma') = 1$ and $W(\sigma\to\sigma') \geq 0$.
In addition, we require the detailed balance condition
\begin{equation}
P(\sigma) W(\sigma\to\sigma') = P(\sigma') W(\sigma'\to\sigma)
\label{eqn:db_eqn}
\end{equation}
to guarantee an equilibrium probability distribution $P(\sigma)$,
i.e.
\begin{equation}
\sum_\sigma P(\sigma) W(\sigma\to\sigma') = P(\sigma').
\end{equation}

It is useful to view the matrix $W(\sigma \to \sigma')$ as composed of two
independent parts --- selection probability $S(\sigma \to \sigma')$ and
acceptance rate $a(\sigma \to \sigma')$.
\begin{equation}
W(\sigma \to \sigma') = S(\sigma \to \sigma') a(\sigma \to \sigma').
\label{eqn:splitW}
\end{equation}
For example, the standard Metropolis algorithm \cite{Metropolis} uses
\begin{equation}
a(\sigma \to \sigma') = \min\left(1, \frac{P(\sigma')}{P(\sigma)}\right),
\qquad\sigma\neq\sigma',
\end{equation}
with $S(\sigma \to \sigma')$ usually set to a constant.
We can easily check that it satisfies the detailed balance condition when
the selection probability $S$ is symmetric, i.e.
$S(\sigma\to\sigma') = S(\sigma'\to\sigma)$.
This symmetry in $S$ can be relaxed for general Monte Carlo simulation
\cite{Hasting} but is needed for the flat histogram sampling algorithm.

By summing up the detailed balance equations for all states $\sigma$ with
energy $E$ and all states $\sigma'$ with energy $E'$, we have
\begin{eqnarray}
\sum_{E(\sigma)=E}\sum_{E(\sigma')=E'} P(\sigma)W(\sigma\to\sigma') =&&\nonumber\\
\sum_{E(\sigma)=E}\sum_{E(\sigma')=E'} P&&(\sigma')W(\sigma'\to\sigma).  
\end{eqnarray}
If the configuration probability distribution is a function of energy only,
i.e. $P(\sigma) \propto f\left(E(\sigma)\right)$, and defining the transition
matrix in energy as
\begin{equation}
T(E\to E') = \frac{1}{n(E)}\sum_{E(\sigma)=E}\sum_{E(\sigma')=E'}
W(\sigma\to\sigma'),
\end{equation}
we have
\begin{equation}
n(E)f(E)T(E\to E') = n(E')f(E')T(E'\to E).
\label{eqn:DBinT}
\end{equation}
The transition matrix $T(E \to E')$ is also a stochastic matrix with the
histogram $h(E)=n(E)f(E)/Z$ being the equilibrium distribution:
\begin{equation}
\sum_E h(E) T(E\to E') = h(E').
\end{equation}

Since the acceptance rate $a(\sigma\to\sigma')$ in Eq.~(\ref{eqn:splitW}) is the
same for configurations with a fixed energy, $T(E \to E')$ can also be written
as a product of two independent factors --- the infinite temperature transition
matrix $T_\infty(E \to E')$ and acceptance rate in energy $a(E \to E')$:
\begin{equation}
T(E\to E') = T_\infty(E \to E')\; a(E\to E'),
\quad E\neq E',
\label{eqn:TNa}
\end{equation}
where $a(E \to E')$ can be any acceptance rate satisfying
Eq.~(\ref{eqn:db_eqn}), the detailed balance equation, e.g. Metropolis
acceptance rate $\min(1, f(E')/f(E))$, and
\begin{equation}
T_\infty(E \to E') =
\sum_{E(\sigma)=E}\sum_{E(\sigma')=E'}\frac{S(\sigma\to\sigma')}{n(E)}.
\label{eqn:defineTi}
\end{equation}

When $S(\sigma \to \sigma')$ is taken to be a constant, the expression can
be simplified. For example, in spin systems we usually pick a spin randomly
to be flipped so that $S(\sigma \to \sigma') = 1/N$, for $\sigma$ and
$\sigma'$ related by a single spin flip and zero otherwise, where $N$ is the
total number of spins. In this case
\begin{equation}
T_\infty(E \to E') = 
\frac{1}{n(E)}\!\sum_{E(\sigma)=E}\!\frac{N(\sigma,\Delta E)}{N} =
\frac{\langle N(\sigma,\Delta E) \rangle_E}{N},
\label{eqn:defineN}
\end{equation}
where $\Delta E=E'-E$ and $N(\sigma, \Delta E)$ represents the number of 
spins that changes the energy of the current configuration $\sigma$ by $\Delta
E$ when flipped, i.e. $N(\sigma, \Delta E)/N = \sum_{E(\sigma') = E'} S(\sigma
\to \sigma')$. 

Substituting the Eq.~(\ref{eqn:TNa}) into Eq.~(\ref{eqn:DBinT}), and
using the equation
\begin{equation}
f(E) \; a(E \to E') = f(E') \; a(E' \to E),
\end{equation}
which is derived from the detailed balance equation Eq.~(\ref{eqn:db_eqn}) and
Eq.~(\ref{eqn:splitW}) with the requirement that $S(\sigma\to\sigma')$ is
symmetric, which also implies that moves must be reversible, we obtained a set
of equations
\begin{equation}
n(E) T_\infty(E \to E') = n(E') T_\infty(E' \to E).
\label{eqn:DBeqn}
\end{equation}
When $S(\sigma \to \sigma')$ is fixed to a constant, we can write
\begin{equation}
 n(E) \langle N(\sigma,\Delta E) \rangle_E =
 n(E') \langle N(\sigma',-\Delta E) \rangle _{E'}.
\label{eqn:broadeqn}
\end{equation}
This equation, known as broad histogram equation, was first presented by
Oliveira in \cite{Oliveira-EurB} and by Berg and Hansmann \cite{Berg-EurB}.
Their derivation is based on property that moves between configurations are
reversible and is somewhat simpler than our arguments above.

However, the interpretation of $N(\sigma, \Delta E)$ cannot be applied when the 
selection probability is not a constant as the definition of $N(\sigma,
\Delta E)$ limits it to be an integer. The general definition of
$T_\infty(E\to E')$ that can be applied to any choice of $S(\sigma\to\sigma')$
is given by Eq.~(\ref{eqn:defineTi}). Eq.~(\ref{eqn:DBeqn}) is a general
equation with two important assumptions made in our formulation: the probability
of every configuration is a function of its energy only and allowable moves
between any pair of configurations are selected with the same probability in
both directions.

The energy detailed balance equation can thus also be written as
\begin{eqnarray}
h(E) \, T_\infty(E \to E') \, a(E\to E') =&&\nonumber\\
h(E')\, T_\infty(E' &&\to E) \, a(E'\to E).
\end{eqnarray}
If we require that the histogram is constant or flat, i.e. $h(E)=h(E')$,
we choose the acceptance rate
\begin{equation}
a(E\to E') = \min\left(1, \frac{T_\infty(E' \to E)}{T_\infty(E \to E')}\right).
\label{eqn:accept_rate}
\end{equation}
This equation was first proposed in \cite{Wang-broad} and \cite{Li-thesis}.
While equivalent to the entropic sampling \cite{Lee} acceptance rate $\min(1,
n(E)/n(E'))$, the simulation procedure is different. We use a cumulative
average to construct the acceptance rate during simulation, i.e.
\begin{equation}
T_\infty(E \to E') \approx
\frac{1}{H(E)} \sum_{i=1}^m \delta_{E(\sigma^i),E}
\!\sum_{E(\sigma'^i)=E'} S(\sigma^i \to \sigma'^i),
\end{equation}
where $H(E) = \sum_{i=1}^m \delta_{E(\sigma^i), E}$ is the cumulative histogram
for a given energy, $m$ is the total number of samples generated so far, 
$\sigma^i$ is the configuration at step $i$ of the simulation and $\sigma'^i$
are the available configurations that can be reached from $\sigma^i$ in one
move. Whenever data is unavailable to compute the acceptance rate,
we simply accept the move in order to sample the unexplored states.

\begin{figure}[t]
\begin{center}
\mbox{\epsfig{file=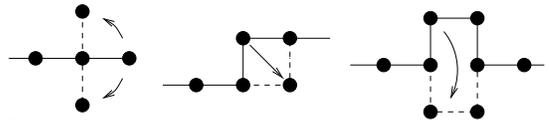, width=0.40\textwidth}}
\caption{Three types of moves, i.e. end, corner and crankshaft move.}
\label{fig:moveset.eps}
\end{center}
\end{figure}

To model protein folding by a Markov process, it is necessary to first define
the move set.  There are several choices and it has been shown that different
move sets can affect kinetic quantities, like the mean first passage time
\cite{Chan-Dill}.
We use a local move set consisting of end, corner and the crankshaft move
\cite{Verdier,Kremer}. These are shown in Fig.~\ref{fig:moveset.eps}.
Note that the end moves are restricted to $90^\circ$
rotations and thus can have one or two possible positions depending on current
configuration. The set of valid moves from current conformation
$\sigma$ to new conformation $\sigma'$ is those that can be performed
using end, corner or crankshaft move, while preserving the excluded volume
constraint.

The choice of the move set used will affect the sampling space of configurations
and also the correlation time of our algorithm. It can be shown that
our set of local moves is non-ergodic, meaning that it does not generate all
possible self-avoiding chains on the lattice. However, the number of 
such configurations is probably negligible compared to our sampling error
\cite{Kremer}. We can also consider the definition of our model to include
only those states accessible by the move set specified. As long as our native
state is accessible, there will not be a problem.

The ergodicity problem can be overcome by pivot moves \cite{Sokal}.
Pivot moves are defined by randomly choosing a monomer as a pivot and rotating
or reflecting one segment of the chain with the pivot as origin. However, pivot
moves also fills more entry in our transition matrix making it less banded.
Moreover, it is also found that pivot moves do not necessary lead to faster
dynamics in the simulation \cite{Clote}.

Although some studies only consider conformations as distinct when they are
not related by reflection or rotational symmetry, the choice should not
matter since they differ by a factor of 8 in counting the number of states.
For the special configurations on a straight line, the factor is 4. However,
such configurations have the highest energy and are thus almost negligible.

While the selection probability is commonly taken to be $1/N$ in spin
systems, we have some freedom in deciding the specific selection probabilities
based on different moves in polymer systems.  We list three possibilities.

\begin{itemize}
\item Given a configuration, we can construct a list of all possible valid
moves satisfying the excluded volume constraint. Each move is selected with a
fixed equal probability. In practice, we assign moves into a list that can
contain up to $M$ moves. Since $M$ is the upper bound to maximum possible valid
moves available to any configuration, the remaining unassigned moves are
considered invalid. A move is picked at random. If it is valid, we accept the
move using the acceptance ratio in Eq.~(\ref{eqn:accept_rate}). In our
simulation, we set $M$ equal to $N$, the number of monomers.

\item Given a configuration, we pick a monomer at random. If it is an end
monomer, it is able to rotate to two possible positions.\footnote{This is
different from our case where $180^\circ$ rotations are not allowed. It is
necessary for the number of possible moves of each type be unambiguous for this
case.}
We select one randomly without considering the excluded volume constraint.
If it is not an end monomer, then corner or crankshaft moves are possible.
These two moves are mutually exclusive.  Once a move is selected, we check the
excluded volume constraint. If a move results in moving a monomer into an
already occupied position, the move is rejected. Otherwise, we accept the move
using the acceptance ratio in Eq.~(\ref{eqn:accept_rate}). For cases
where no moves are possible for the monomer picked, the current configuration is
included in the average and the time step is incremented by one. This method of
selecting move is used in Ref.~\cite{Socci}. It relies on the fact that in
moving to a new configuration using a particular type of move, only the same
type of move in reverse can bring it back to the original configuration. Thus,
the move set must be chosen with this property to ensure that selection
probability is symmetric.

\item We can designate end moves and corner moves as \textsl{1-monomer move}
and crankshaft move as \textsl{2-monomer move}. With probability e.g.
0.2, we choose to perform 1-monomer move; if an end monomer is picked,
end moves are selected and corner moves otherwise, in the same manner as above
(see footnote).
With remaining probability 0.8, pick a monomer from 1 to $N-3$, a 2-monomer
move is selected if monomers from $i$ to $i+3$ forms a crankshaft otherwise
a move is unsuccessful. This method is used in Ref.~\cite{Sali}.
\end{itemize}

For compact configurations, the rejection rate for the second method would
be very high. We use the first method, although a large $M$
can also make the method inefficient for compact (thus low-energy)
configurations since the number of possible moves is low compared to the value
of $M$. However, this can be overcome if an N-fold way simulation \cite{Bortz}
is done. A move is always accepted in the N-fold way and the average lifetime
of a configuration is taken into account when averaging. The first method also
has the advantage of saving some computations. Since the selection probability
is a constant, we can just add a constant (equivalent to counting moves) when
calculating $T_\infty(E \to E')$ in Eq.~(\ref{eqn:defineTi}).
The list of moves can also be used in constructing an N-fold way simulation.
For other choices of selection probability, we will have to calculate
$S(\sigma \to \sigma')$ explicitly for each move before adding. Once
$T_\infty(E \to E')$ is sampled, we solve Eq.~(\ref{eqn:DBeqn}). Since $n(E)$
varies by a huge order of magnitude, we solved for $\ln n(E)$ instead.  Broad
histogram method uses a forward difference scheme of integration. We solve it
instead using a least squares method. When multiple simulations are performed,
we can view it as an optimization problem taking the variance of sampling data
into account \cite{Wang-tmmc}.

\section{Numerical Results}

We present results on a sequence with 14 monomers with the sequence
$\mathsf{HHHPHPHPPHPHPH}$. Through enumeration, we found that this
sequence have a unique ground state (i.e. 8 possible configurations in our
counting) of energy $-7$. The full density of states is given in
Table~\ref{tbl:dos}.

\begin{figure}[tbh]
\centering
\epsfig{file=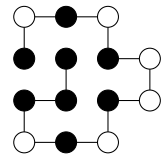, width=2cm}
\caption{Native state for the sequence $\mathsf{HHHPHPHPPHPHPH}$.}
\label{fig:S14.eps}
\end{figure}

\begin{table}[hbt]
\caption{Density of states for the sequence $\mathsf{HHHPHPHPPHPHPH}$. 
$\tilde n(E)$ is Monte Carlo results using flat histogram sampling
algorithm and $n(E)$ is through enumeration.}
\label{tbl:dos}
\centering
\begin{tabular}{rrrr}
$E$ & $n(E)$ & $\tilde n(E)$ & \% error \\ \tableline
 0 & 581340 & 540416.37 & 7.04 \\ 
-1 & 228416 & 217016.11 & 5.00 \\ 
-2 & 56344  & 55837.55  & 0.89 \\ 
-3 & 12472  & 12666.23  & 1.56 \\ 
-4 & 2432   & 2465.45   & 1.38 \\ 
-5 & 464    & 485.93    & 4.73 \\ 
-6 & 24     & 23.66     & 1.42 \\ 
-7 & 8      & 8         & 0    \\
\end{tabular}
\end{table}

\begin{figure}[tbh]
\begin{center}
\mbox{\epsfig{file=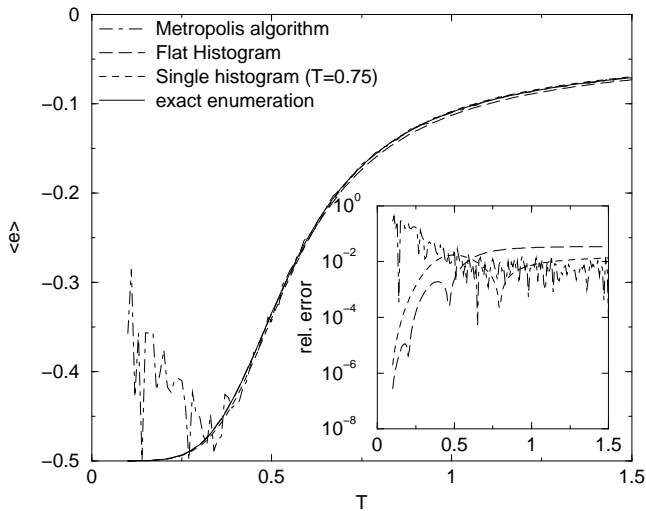, width=0.48\textwidth}}
\caption{The average energy per monomer against temperature. The insert
shows the relative error between using enumerated $n(E)$ and each method.}
\label{fig:S14e0.eps}
\end{center}
\end{figure}

\begin{figure}[htb]
\begin{center}
\mbox{\epsfig{file=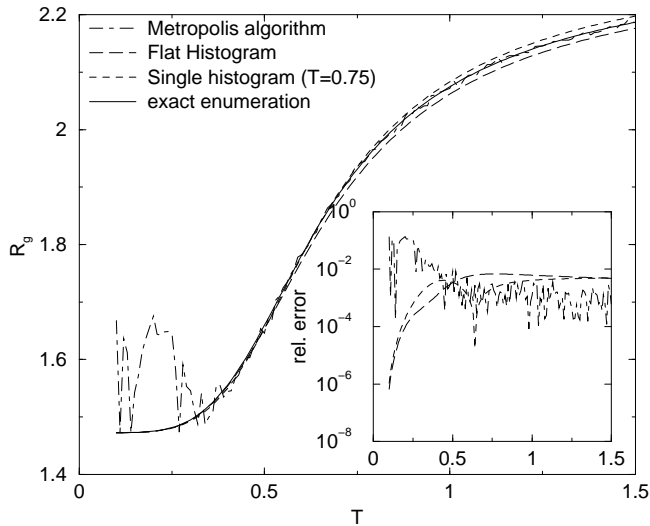, width=0.48\textwidth}}
\caption{The radius of gyration against temperature. The insert
shows the relative error between using enumerated $n(E)$ and each method.}
\label{fig:S14r0.eps}
\end{center}
\end{figure}

The average energy per monomer and radius of gyration are plotted in
Fig.~\ref{fig:S14e0.eps} and \ref{fig:S14r0.eps} and are compared
with the single histogram method and the Metropolis algorithm. We used $10^6$
Monte Carlo steps for the flat histogram sampling algorithm and each temperature
point of Metropolis algorithm and also the single histogram reweighted at
$T=0.75$. There are about 140 temperature points in the Metropolis simulation,
which requires around a 100 times more computing time compared to the flat
histogram simulation.

\begin{figure}[htb]
\begin{center}
\mbox{\epsfig{file=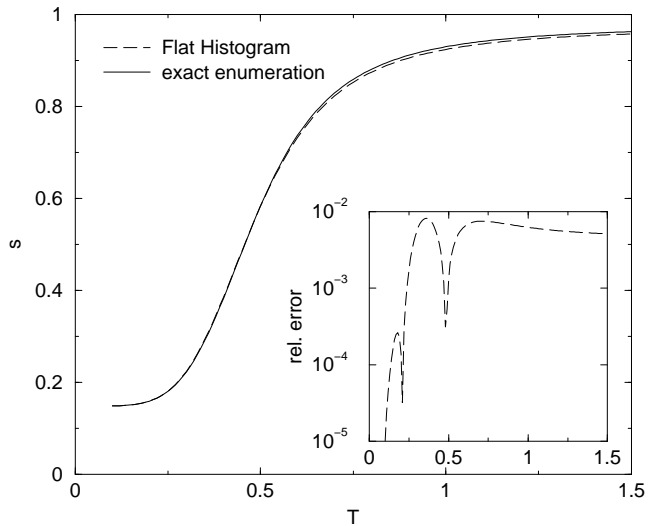, width=0.48\textwidth}}
\caption{The average entropy per monomer against temperature. The insert
shows the relative error.}
\label{fig:S14s0.eps}
\end{center}
\end{figure}

The plots indicate that the Metropolis algorithm becomes
unreliable below around $T=0.5$. This slowing down in dynamics is also
found in \cite{Socci} and attributed to the increasingly deep kinetic traps
with decreasing temperature. The flat histogram sampling algorithm is
unaffected by this effect with slightly better accuracy for low temperature
range.  The single histogram method also produces roughly the same degree of
accuracy. Here we do not observe the reweighting errors due to the
exponential decay of the canonical distribution because the energy spectrum
is narrow and thus adequately sampled. The single histogram method
works well only in such a situation. The entropy which can be easily calculated
from the knowledge of $n(E)$, is shown in Fig.~\ref{fig:S14s0.eps}. It is 
difficult to obtain this from Metropolis simulation.

Finding the energy of the native state is also an important task in protein
folding. To generate the native states using Metropolis algorithm, the
temperature must be low enough so that the canonical distribution covers
the low energy range adequately. However, the canonical distribution has
width $\sqrt N$ and low temperature simulation causes the system to be trapped
in local minima. Various methods, such as genetic algorithm \cite{Unger} and
methods employing heuristic \cite{Toma} have been proposed to overcome this
problem. Often, an annealing schedule is adopted whereby the temperature is
lowered as the simulation progress. There is no standard way and considerable
trial-and-error is necessary.

The flat histogram sampling algorithm can be used as a method for determining
the native state energy. Since the energy barriers no longer exist, we expect a
random walk along the energy scale in the ideal case. An advantage is
that in most polymer models, the energy range do not increase rapidly with
system size. We also do not have to devise any annealing schedule or adjust
many parameters. We can therefore use our algorithm for determining the native
state energy. Although we cannot attach any physical significance to the time
for finding the native state since the ensemble is non-Boltzmann
(multicanonical), it is still useful for analyzing the performance of our
algorithm. This native state time $\tau_0$, the time to reach the native state
from an unfolded conformation, is shown for four sequences in
Table~\ref{tbl:groundstate_time}. We select the first two sequence to have
unique native state. The other two sequences were taken from \cite{Unger}.

\begin{table}
\caption{Sequence of HP monomers used in simulation with their lowest energy
and native state time averaged from $10^4$ simulations. The last column is the
standard deviation of $\tau_0$.}
\label{tbl:groundstate_time}
\centering
\begin{tabular}{rrrrr}
$N$ & Sequence & $E_0$ & $\tau_0$ & $\sigma_{\tau_0}$ \\ \tableline
10  & \scriptsize{\sffamily HPHPPHPPHH} & -4 & 339.7 & 405.7\\
14  & \scriptsize{\sffamily HHHPHPHPPHPHPH} & -7 & 5641.4 & 6340.4\\
20  & \scriptsize{\sffamily HPHPPHHPHPPHPHHPPHPH} & -9 & 81788.1 & 85478.5 \\
25  & \scriptsize{\sffamily PPHPPHHPPPPHHPPPPHHPPPPHH} & -8 & 196137.9 & 318722.3\\
\end{tabular}
\end{table}

The tunnelling time can also be used as a measure of the efficiency of our
algorithm. We denote $\tau_u$, the ``up'' tunnelling time as the average MCS
taken for a state with minimum energy to reach a state with maximum energy
while $\tau_d$, the ``down'' tunnelling time is for the opposite direction.
These are shown in Table~\ref{tbl:tunnel_time}. Unlike spin systems, where
the two tunnelling times are the same due to the symmetry in the Hamiltonian,
it is faster to tunnel to higher energies than to lower energies.

\begin{figure}[htb]
\begin{center}
\mbox{\epsfig{file=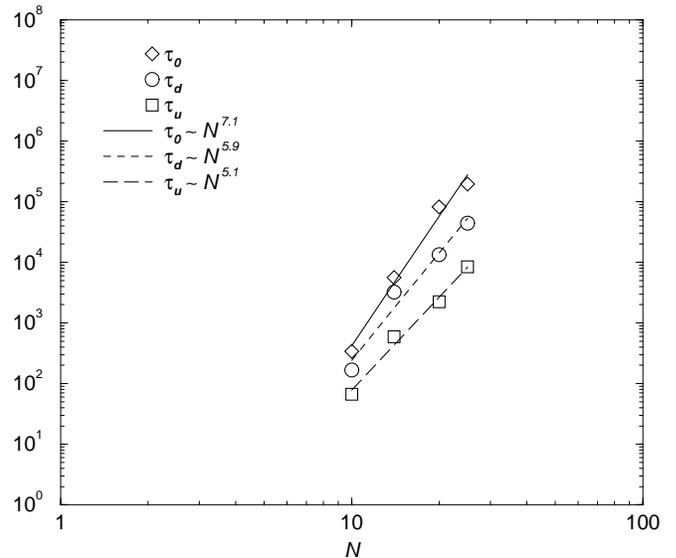, width=0.48\textwidth}}
\caption{The points are native state time and tunnelling times obtained from
simulation. The straight lines are fits to a power law $\tau \propto N^p$.}
\label{fig:tunnel1.eps}
\end{center}
\end{figure}

Fig.~\ref{fig:tunnel1.eps} shows the general behaviour of the native state time
and tunnelling times as the size increases. It is usual in disorder systems
to average over different random realizations corresponding to different sets
of coupling constants or sequences. However, it is recognized that proteins are
not random sequences since they fold into unique native states with
specific properties. The probability of selecting a sequence with this property
from all possible sequences is very small. This leads to the problem of
designing sequences with protein-like properties.

\begin{table}[ht]
\caption{Tunnelling times, $\tau_u$ and $\tau_d$ for different sequences.}
\label{tbl:tunnel_time}
\centering
\begin{tabular}{rrrrrrrr}
$N$ & $E_0$ & $\tau_u$ &  std & count & $\tau_d$ & std & count \\ \tableline
 10 &  -4 &      66.8 &   60.7 & 4272 &     167.3 &   208.3 & 4271\\
 14 &  -7 &     592.6 &  551.5 & 2615 &    3229.3 &  3940.6 & 2615\\
 20 &  -9 &    2224.3 & 3612.3 &  638 &   13427.8 & 13976.7 &  638\\
 25 &  -8 &    8418.0 & 9616.3 &  190 &   44214.1 & 61316.7 &  190\\
\end{tabular}
\end{table}

We fit the times to find the minimum energy (native) states and the tunnelling
times for the four sequences to a power law. The time to find native state
follows approximately $\tau_0 \propto N^{7.1}$.
The ``up'' and ``down'' tunnelling time gives the fitted parameter 
$\tau_u \propto N^{5.1}$ and $\tau_d \propto N^{5.9}$ respectively.
This suggests that the algorithm takes increasingly longer time to reach
the lowest energy states but moves easily towards the upper energy
levels. It reflects our observation that the flat histogram sampling algorithm
does not scale very well for longer chains especially when the density of
states increase sharply with energy. This also implies that the performance is
not ideal for sequences with a unique native ground state. We note that our
simulation is non-Markovian and thus the convergence is difficult to analyze.
This can lead to detailed balance violation \cite{Wang-Lee}. However, this
problem can be alleviated by a two pass simulation. The first pass is the same
as before. The second pass uses a fixed flip rate, $\min\left(1,
n(E)/n(E')\right)$, obtained from the first pass.

\section{Conclusion}

We have shown that the flat histogram sampling algorithm, which was first
proposed and implemented for spin systems, can be used for the simulation of
lattice polymer systems. We give some measures of its accuracy and also
efficiency in terms of thermodynamics properties, native state time and
tunnelling times. 
The current implementation is useful for up to about 20 monomer HP chain. 
However, we would like to emphasize that the flat histogram sampling algorithm
is still vastly superior to Metropolis algorithm, especially in terms of
accuracy for low temperature properties. The simulation time is modest as there
is no need for simulation at each temperature point. It also has the advantage
of easily obtaining the density of states for free energy or entropy
calculations.  While the algorithm is rather basic, there are improvements to
be made such as the extensions and modifications proposed in
Ref.~\cite{Wang-tmmc}.


\begin{thebibliography}{99}
\bibitem{Binder} K. Binder (ed), \textsl{The Monte Carlo Method in Condensed
	Matter Physics}, Topics in Applied Phys, Vol 71, 2nd ed,
	(Springer-Verlag, 1995).

\bibitem{Metropolis} N. Metropolis, A. W. Rosenbluth, M. N. Rosenbluth,
	A. H. Teller and E. Teller, \textsl{J. Chem. Phys.} {\bf 21},
	1087 (1953).

\bibitem{Ferrenberg} A. M. Ferrenberg and R. H. Swendsen, \textsl{Phys. Rev.
	Lett.} {\bf 61}, 2635 (1988); A. M. Ferrenberg and R. H. Swendsen,
	\textsl{Phys. Rev. Lett.} {\bf 63}, 1195 and 1658 (1989).

\bibitem{Berg} B. A. Berg and T. Neuhaus, \textsl{Phys. Lett.} {\bf B 267},
	249 (1991), B. A. Berg and T. Celik, \textsl{Phys. Rev. Lett.} {\bf 69},
	2292 (1992).

\bibitem{Lee} J. Lee, \textsl{Phys. Rev. Lett.} {\bf 71}, 211 (1993).

\bibitem{Wang-broad} J.-S. Wang, \textsl{Eur. Phys. J. B} {\bf 8}, 287 (1999).

\bibitem{Wang-tmc} J.-S. Wang, T. K. Tay, and R. H. Swendsen, \textsl{Phys.
	Rev. Lett.} {\bf 82}, 476 (1999).

\bibitem{Wang-tmmc} J.-S. Wang and R. H. Swendsen, \textsl{Transition Matrix
	Monte Carlo Method}, preprint (cond-mat/0104418).

\bibitem{Li-thesis} S.-T. Li, \textsl ``The transition matrix Monte Carlo
	method", Ph.D. dissertation, Carnegie Mellon University (1999),
	unpublished.

\bibitem{Oliveira} P. M. C. de Oliveira, T. J. P. Penna and H. J. Herrmann,
	\emph{Braz. J. Phys.} \textbf{26}, 677 (1996); P. M. C. de Oliveira,
	\emph{Braz. J. Phys.} \textbf{30}, 195 (2000).

\bibitem{Oliveira-EurB} P. M. C. de Oliveira, T. J. P Penna and H. J. Herrmann,
	\textsl{Eur. Phys. J. B} {\bf 1}, 205 (1998).

\bibitem{Oliveira-exact} P. M. C. de Oliveira, \textsl{Eur. Phys. J. B}
	{\bf 6}, 111 (1998).

\bibitem{Lau-HP} K. F. Lau and K. A. Dill, \textsl{Macromolecules}
	{\bf 22}, 3986 (1989).

\bibitem{Dill-review} K. A. Dill, S. Bromberg, K. Yue, K. M. Fiebig, D. P. Yee,
	P. D. Thomas and H. S. Chan, \textsl{Protein Sci.} {\bf 4}, 561
	(1995).

\bibitem{Clote} P. Clote and R. Backofen, \textsl{Computational Molecular
	Biology} p229-232, (John Wiley, 2000).

\bibitem{Yue} K. Yue, K. M. Fiebig, P. D. Thomas, H. S. Chan and E. I.
	Shakhnovich and K. A. Dill, \textsl{Proc. Natl. Acad. Sci. USA}
	{\bf 92}, 325 (1995).

\bibitem{Hasting} W. K. Hastings, \textsl{Biometrika} {\bf 57}, 97 (1970).

\bibitem{Berg-EurB} B. A. Berg and U. H. E. Hansmann, \textsl{Eur. Phys. J. B}
	{\bf 6}, 395 (1998).

\bibitem{Chan-Dill} H. S. Chan and K. A. Dill, \textsl{J. Chem Phys.}
	{\bf 99}, 2116 (1993) and {\bf 100}, 9238 (1994).

\bibitem{Verdier} P. H. Verdier and W. H. Stockmayer, \textsl{J. Chem Phys.}
	{\bf 36}, 227 (1962).

\bibitem{Kremer} K. Kremer and K. Binder, \textsl{Comp. Phys. Rept.} {\bf 7},
	259 (1988).

\bibitem{Sokal} N. Madras and A. D. Sokal, \textsl{J. Stat. Phys.} {\bf 50},
	109 (1988).

\bibitem{Socci} N. D. Socci and J. N. Onuchic, \textsl{J. Chem. Phys.}
	{\bf 101}, 1519 (1994) and {\bf 103}, 4732 (1995).

\bibitem{Sali} A. Sali, E. I. Shakhnovich and M. Karplus, \textsl{J. Mol. Biol.}
	{\bf 235}, 1614 (1994).

\bibitem{Bortz} A. B. Bortz, M. H. Kalos and J. L. Lebowitz,
	\textsl{J. Comput. Phys.} {\bf 17}, 10 (1975).

\bibitem{Unger} R. Unger and J. Moult, \textsl{J. Mol. Biol.} {\bf 231}, 75
	(1993).

\bibitem{Toma} L. Toma and S. Toma, \textsl{Protein Sci.} {\bf 5}, 147 (1996).

\bibitem{Wang-Lee} J.-S. Wang and L. W. Lee, \textsl{Comp. Phys. Commun.}
	{\bf 127}, 131 (2000).
\end{thebibliography}
\end{document}